\documentclass[sigconf, nonacm]{acmart}
\AtBeginDocument{
  \providecommand\BibTeX{{
    \normalfont B\kern-0.5em{\scshape i\kern-0.25em b}\kern-0.8em\TeX}}}

\usepackage{subcaption}

\begin{document}

\title{Virtual Buddy: Redefining Conversational AI Interactions for Individuals with Hand Motor Disabilities}

\author{Atieh Taheri}
\affiliation{
  \institution{University of California}
  \city{Santa Barbara}
  \state{CA}
  \country{USA}
  }
\email{a_taheri@ucsb.edu}

\author{Purav Bhardwaj}
\affiliation{
  \institution{University of California}
  \city{Santa Barbara}
  \state{CA}
  \country{USA}
  }
\email{purav@ucsb.edu}

\author{Arthur Caetano}
\affiliation{
  \institution{University of California}
  \city{Santa Barbara}
  \state{CA}
  \country{USA}
  }
\email{caetano@cs.ucsb.edu}

\author{Alice Zhong}
\affiliation{
  \institution{University of California}
  \city{Santa Barbara}
  \state{CA}
  \country{USA}
  }
\email{alicezhong@ucsb.edu}

\author{Misha Sra}
\affiliation{%
  \institution{University of California}
  \city{Santa Barbara}
  \state{CA}
  \country{USA}
  }
\email{sra@ucsb.edu} 

\renewcommand{\shortauthors}{Taheri, et al.}

\definecolor{mygreen}{RGB}{0,120,0}

\newcommand\aticomment[1]{\textcolor{purple}{At: #1}}
\newcommand\purcomment[1]{\textcolor{blue}{Purav: #1}}
\newcommand\arthcomment[1]{\textcolor{mygreen}{Arthur:#1}}
\newcommand\alicomment[1]{\textcolor{orange}{Alice: #1}}
\newcommand\mcomment[1]{\textcolor{red}{Misha: #1}}

\begin{abstract}
    Advances in artificial intelligence have transformed the paradigm of human-computer interaction, with the development of conversational AI systems playing a pivotal role. These systems employ technologies such as natural language processing and machine learning to simulate intelligent and human-like conversations. Driven by the personal experience of an individual with a neuromuscular disease who faces challenges with leaving home and contends with limited hand-motor control when operating digital systems, including conversational AI platforms, we propose a method aimed at enriching their interaction with conversational AI. Our prototype allows the creation of multiple agent personas based on hobbies and interests, to support topic-based conversations. In contrast with existing systems, such as Replika, that offer a 1:1 relation with a virtual agent, our design enables one-to-many relationships, easing the process of interaction for this individual by reducing the need for constant data input. We can imagine our prototype potentially helping others who are in a similar situation with reduced typing/input ability.
    
\end{abstract}

\ccsdesc[500]{Human-centered computing~Human computer interaction (HCI)}
\ccsdesc[500]{Human-centered computing~Accessibility technologies}

\keywords{conversational AI, hand motor disability, user-centric design, accessibility, virtual personas}

\received{20 February 2007}
\received[revised]{12 March 2009}
\received[accepted]{5 June 2009}

\maketitle

\section{Introduction}
Conversational AI is fast proving to be a valuable tool in our day-to-day lives, providing support, companionship, and stress relief~\cite{sharma2023human, meng2021emotional, qi2021conversational, medeiros2021can}. It is now plausible to consider the use of this technology as a means of mitigating social isolation~\cite{gadbois2022findings, soraa2020mitigating, de2020effectiveness, shum2018eliza}.
However, current conversational AI technologies have largely catered to the general audience. The unique needs of people with mobility disabilities, who may not be able to leave home to engage in social interactions, or those who have motor impairments, that make it challenging to interact with computers, have not been the primary focus of their design, despite clear benefits. Unlike mobility disability, which limits movement, motor disability affects muscle control and impacts mobility, among other things. Our work focuses on people with motor challenges, addressing the physical barriers that impact interactions with conversational AI as a socialization aid.

The motivation and design decisions for this work originate from the firsthand experiences of the primary author, who lives with Spinal Muscular Atrophy (SMA), a severe motor neuron disease. She uses a wheelchair for mobility and her right thumb for all computer interactions. Although AI-powered real-time conversations and question-answering offer substantial potential, their accessibility is limited.

For instance, Taheri has a deep interest in learning a new language. Attending in-person Italian lessons is difficult for her. Interacting with a conversational AI could provide a dynamic and real-time experience similar to talking and learning from a language tutor but one that tirelessly provides explanations and practice, is available any time of day, is not impacted by personal circumstances and moods, and in general, adjusts to her needs. However, existing conversational AI systems constantly require manual information input to make the agent act in a desired role, making them quickly unusable for Taheri, as input with one thumb is time-consuming and exhausting. This inspired us to focus on the potential of conversational AI for her and for individuals in similar situations, by smoothing out the onboarding process of engaging with chatbots by minimizing user input.

In this work, we introduce a prototype called Virtual Buddy, also referred to as V-Buddy, that focuses on reducing friction in the onboarding process of interaction with conversational agents.
V-Buddy offers the ability to create multiple personas, optionally via pre-filled templates with easy-to-select options, each giving the AI agent a different role and personality, to assist focused or topic-relevant conversations. The personas created on V-Buddy remain usable across interaction sessions without requiring repeated or further information from the user over time. However, if the user so chooses, they are editable allowing the user to modify the persona as needed. With multiple available personas, V-Buddy also enables supporting one-to-many relationships (e.g., one user can have many different AI buddies in the same app) as opposed to the one-to-one relationship the current conversational AI systems offer. 

\section{Background}
Conversational AI agents with their unique design focus and user experiences are now available on multiple platforms. Highlighting some of these and their intended purposes, we aim to pinpoint their limitations in assisting users with hand-motor disabilities.
Replika~\cite{replikaReplika}, designed as a general companion, provides emotional support and companionship to its users. It has the distinct ability to adapt to a user's personality and preferences, carving a unique relationship with each user. However, it is only accessible via mobile devices, which is inaccessible to Taheri and potentially to others with hand-motor impairments. Caryn AI~\cite{carynCarynai}, still in beta, allows users to converse with a virtual influencer. Although its unique training method using GPT-4~\cite{OpenAI2023-cl} and Forever Voices~\cite{ForeverV41:online} suggests potential, its performance is still being evaluated. 
These systems, and others like them, primarily foster a one-to-one relationship with the user, useful for entertainment and skill-building. To allow the agent to adopt a different persona, complete with unique behaviors and attitudes, the user must provide the necessary information repeatedly. This requires a significant amount of input effort, difficult for someone unable to type. Our proposed system aims to overcome this limitation with pre-created personas to facilitate building one-to-many relationships with different AI agents.  

\section{V-Buddy}
\begin{figure}[t]
    \begin{subfigure}[b]{0.45\columnwidth}
    \centering
\includegraphics[height=\textwidth]{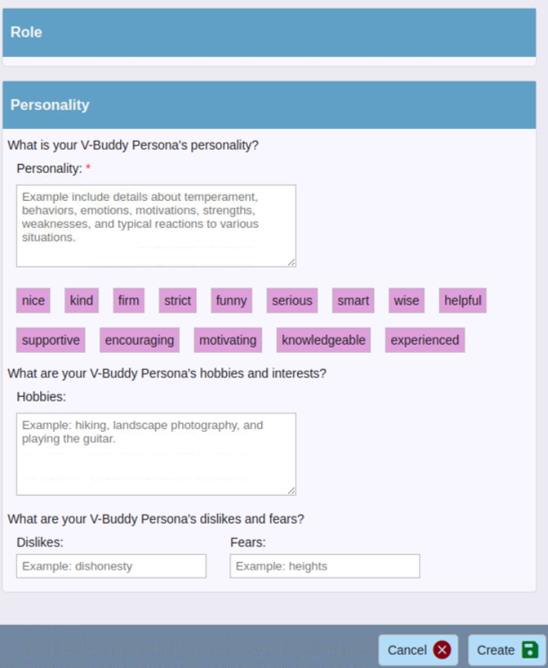}
    \end{subfigure}
    \begin{subfigure}[b]{0.45\columnwidth}
    \centering
\includegraphics[height=\textwidth]{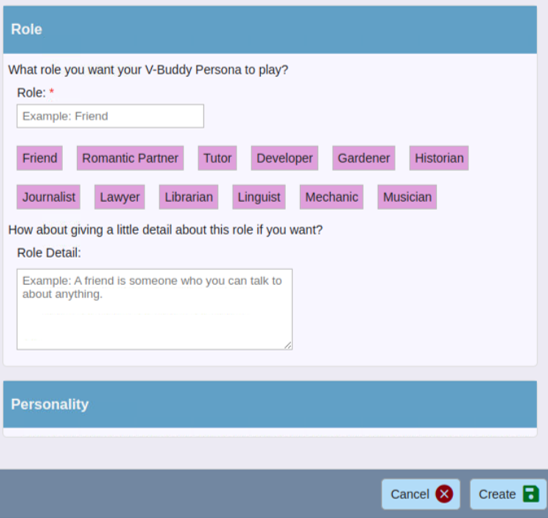}
    \end{subfigure}
    \caption{Screenshots of the interface, showing easy-to-select personality (e.g., kind, strict) and role (e.g., Gardener, Musician) options to generate a V-Buddy; customizable as needed.}
    \label{fig:app}
\end{figure}

V-Buddy is a web-based system that enables users to create several virtual personas facilitating one-to-many interactions focusing on specific hobbies and personality types of agents. The \emph{User} component of V-Buddy collects basic details like the user's name, pronouns, and optionally, their general needs, including disability-related ones. The \emph{V-Buddy Personas} component lets users ``create'' a new persona or ``select'' from their previously created ones. Users can customize their persona based on three attributes: Role, Personality, and Needs, each offering pre-filled options for ease. For example, Role can be a friend or tutor, Personality can be kind or witty, and Needs might incorporate disability accommodations. Figure~\ref{fig:app} shows the interface for setting the role and personality. Once set, these preferences are stored in a MongoDB-based~\cite{mongodbMongoDBDeveloper} database. Through the \emph{Converse} component, users can begin to interact with their created persona, with responses powered by GPT-4 ~\cite{OpenAI2023-cl}. V-Buddy's GUI adheres to WCAG's operability principles, offering tailored options for users with limited hand-motor control. It directs GPT-4 with specific prompts to align responses with user-selected roles and personalities, such as guiding a ``friend'' role with a ``witty'' personality to elicit friendly and humorous conversations.

V-Buddy is an early proof-of-concept, designed with the goal of adapting conversational AI to be accessible to Taheri and potentially, people in a similar situation, where typing is challenging or not possible due to temporary injury or long term disability. While speech-based input is an option, it may not always work due to varying accents, levels of speech clarity, and tongue muscle control and hence not considered in our current prototype.
Including specific needs for both the user and the virtual buddy in the design process can lead to more empathic and understanding interactions, which can potentially strengthen the bond between the user and the virtual buddy~\cite{bennett2019promise, zhu2023toward}.
V-Buddy offers easy-to-select options, generated using a GPT-4 model inspired by Valencia et al.'s work~\cite{valencia2023less}, which showcased 
the utility of LLMs to expand short text snippets into longer output aiding augmentative and alternative communication (AAC) for non-verbal individuals. The options provided are contextually appropriate to the ongoing conversation, aiming to improve user experience and reduce physical strain. 

\subsection{Limitations and Future Work}
V-Buddy has limitations that need to be acknowledged and will be resolved in a future version.
The first one lies in balancing ease of use with user agency. While offering selectable options would ease interaction and reduce physical burden on Taheri, it may limit user autonomy in shaping their virtual buddy. We attempt to resolve this by offering a vast array of options, and also provide a way for the user to input their own descriptors if they so choose. A future user evaluation would help us better understand if and how these two options serve the needs of users who have challenges interacting with a computer. 
The second risk and one that is true of any conversational AI system, is that of users becoming overly attached to their chatbots. Prior studies have indicated that an overly human-like quality in AI companions can lead to negative interaction outcomes~\cite{park2023human}, such as harmful dependencies and unrealistic expectations, potentially causing psychological issues~\cite{possati2022psychoanalyzing, boine2023emotional}. Building such tools ethically and responsibly needs careful evaluations, both short and long-term, to understand how the designs should be manipulated (e.g., make them look less human) to mitigate unintended negative consequences.
V-Buddy is designed to complement, not to replace, human interactions, particularly for those who are unable to frequently engage in in-person social activities.
It can enrich existing social channels and provides extra support for those struggling with currently available conversational AI systems.

\bibliographystyle{ACM-Reference-Format}
\bibliography{arxiv}

\end{document}